# A Continuous Action Space Tree search for INverse desiGn (CASTING) Framework for Materials Discovery

Suvo Banik[1, 2], Troy Loefller[1, 2], Sukriti Manna[1, 2], Srilok Srinivasan[1], Pierre Darancet[1], Henry Chan[1, 2], Alexander Hexemer[3], Subramanian KRS Sankaranarayanan*[1,2]

*skrssank@uic.edu

[1] Center for Nanoscale Materials, Argonne National Laboratory, Lemont, Illinois 60439.
[2] Department of Mechanical and Industrial Engineering, University of Illinois, Chicago, Illinois 60607.
[3] Advanced Light Source (ALS) Division, Lawrence Berkeley National Laboratory, Berkeley, CA 94720.

**Abstract**

Fast and accurate prediction of optimal crystal structure, topology, and microstructures is important for accelerating the design and discovery of new materials for energy applications. As material properties are strongly correlated to the underlying structure and topology, inverse design is emerging as a powerful tool to discover new and increasingly complex materials that meet targeted functionalities. A challenge lies in the exorbitantly large structural and compositional space presented by the various elements and their combinations. Despite the large number of new phases already being populated in the myriads of materials databases, structural prediction is still an active research area given the vast configurational and chemical spaces available for exploration. Speed, accuracy, and scalability are three desirables for any inverse design tool to sample efficiently across such a vast space. Global optimization strategies developed to tackle this challenge include, most notably, evolutionary approaches using genetic algorithms and those based on random sampling. While these approaches have demonstrated the ability to predict new crystal structures that can be used as super-hard materials, semiconductors, and photovoltaic materials to name a few, it is highly desirable to develop approaches that converge faster to the solution, have better solution quality, and are scalable to high dimensionality. Reinforcement learning (RL) approaches are emerging as powerful design tools capable of addressing these issues but primarily operate in discrete action space. As such, their applications to inverse materials design problems are limited owing to the continuous nature of materials search spaces. In this work, we introduce CASTING, which is an RL-based scalable framework for crystal structure, topology, and potentially microstructure prediction. CASTING employs an RL-based continuous search space decision tree (MCTS -Monte Carlo Tree Search) algorithm with three important modifications (i) a modified rewards scheme for improved search space exploration (ii) a "windowing" or "funneling" scheme for improved exploitation and (iii) adaptive sampling during playouts for efficient and scalable search. Using a set of representative examples ranging from metals such as Ag to covalent systems such as C and multicomponent systems (graphane, boron nitride, and complex correlated oxides), we demonstrate the accuracy, the speed of convergence, and the scalability of CASTING to discover new metastable crystal structures and phases that meet the target objective.

**Introduction:**

The properties (chemical, physical, thermal, optical, mechanical to name a few) of any material are intimately tied to its crystal structure, topology and/or microstructure. Design and discovery of crystalline polymorphs *i.e.*, metastable states that are radically different in their structural features, and establishing a correspondence between their structure and performance in a targeted application domain have been a long-standing challenge in the area of material design and synthesis[1, 2]. Crystal Structure Prediction (CSP)[1, 3, 4, 5, 6, 7, 8, 9, 10, 11, 12] involves navigation through a vast configurational and compositional space with high permutational variability which makes the search problem challenging. Global optimization techniques have been traditionally employed to search through the vast configurational and compositional landscape to predict optimal materials for inverse design applications[6, 9, 10, 13, 14, 15, 16]. Alternate approaches were intuition-based and relied on empirical schemes [17]. This not only limits the tractability of the problem but is also very much restrictive in terms of exploration.

In the past few decades, significant algorithmic development[4] and implementation, particularly, in Crystal Structure Prediction (CSP), has unraveled a new paradigm predicting new materials that display exotic properties[2, 4, 5, 18]. Data-driven approach[3, 7], simulated annealing[6, 14], minima hopping[19], and meta dynamics[20, 21] have been used with some success For systems with smaller sizes, even random sampling followed by atomistic relaxation produce structures with stable configurations[22, 23]. Metaheuristic techniques such as evolutionary algorithm[5, 9, 13], particle swarm[10, 15, 16], and basin hopping[24, 25], have subsequently been developed and applied to a multifarious class of materials. This allowed a search for the ground state structures based on the chemical composition and the external conditions. Not only have the crystal structure prediction methods predicated new materials but many of these theoretically predicted configurations have been experimentally synthesized, bridging theory and experiment in design and discovery[26, 27, 28, 29]. More recently, artificial intelligence (AI) and Machine Learning (ML) techniques are emerging as efficient tools in mapping quantitative structure to property relationship (QSPR)[30, 31, 32, 33, 34].

The overall success of any crystal structure or topology prediction methodology is widely dependent on the exploratory nature and the convergence strength of the search algorithm. As the dimensionality of the search space increases with either an increase in size or composition, navigating efficiently through the search space with multiple local minima becomes very challenging[15]. It is also to be noted that more accurate methods like density functional theory (DFT)[35, 36] are computationally expensive and successful implementation of this method in CSP necessitates the algorithm to fasten convergence. In this regard, ML has again led to advances in the development of cheaper surrogate models to represent the underlying materials' physics and chemistry[37, 38]. Furthermore, we note that inverse design involves search across energy surface[2] and configurational space which are continuous in nature. At each of the valleys of this surface lie local minima (metastable states) representing a crystal structure that can map to an exotic property Fig. 1(a). Our target solution represents one such minima as highlighted by point C in Fig. 1(a). The hills in the energy (and corresponding configurational) surface are barriers that can be overcome with suitable thermodynamic conditions such as, for example, temperature, pressure, composition, or their combinations. While we are interested in reaching the target solution, it is also desirable to explore the local minima or metastable states as well. This requires our search algorithm to establish a balance between exploration vs. exploitation. All sampling techniques[13, 16] ,especially in high dimensional space suffer from poor solution quality and/or sluggish convergence owing to an exponentially increasing volume and the number of local minima's[1]. A possible solution to this problem is learning from the explored part of the search space and utilizing the knowledge for further exploration. This can greatly improve the efficiency of the optimizer algorithm being used. Reinforcement learning (RL) with the ability to learn on the fly from the current state of the system and make decisions not only based on the current state of the system but also on the history can greatly aid in overcoming the so-called "curse of dimensionality".

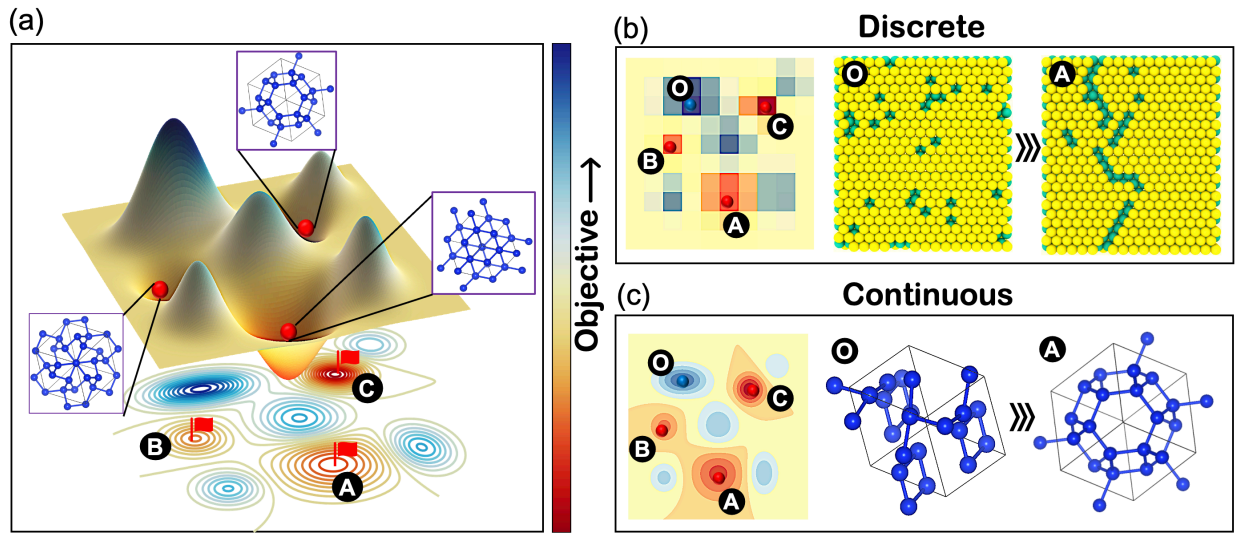

*Fig. 1.* *Schematic illustration of the nature of the search space (discrete vs. continuous) in materials applications. (a) depicts the typical potential energy landscape of materials with different metastable polymorphs at local minima. (b) depicts a discrete action space with defects movement as actions[32]. Moving*

*from a high energy defect configuration situated at O to a relatively stable polymorph at A consists of finite movement in the defect configurational space where navigating the PES involves finite steps with a discrete jump in energy. (c) Crystal structure optimization represents a continuous action or search space problem with infinite possibilities of moving an atom giving a large number of pathways to navigate from high energy polymorph at O to local minima at A.*

RL-based approaches have achieved remarkable success in solving problems with seemingly large intractable search space, such as board games Chess, Shogi, and Go [39, 40], and more recently in materials applications such as chemical synthesis planning[41] or drug discovery[42, 43, 44]. Most of the materials applications to this day have been limited to discrete action spaces[45] Fig. (1)(b), including, for example, optimization of the geometry of lattice defect[31, 32] described as a set of discrete positions on a finite lattice. However, many real-world problems including several grand challenges in materials discovery and design involve decision-making and search in a rather continuous action space[37] Fig. 1(c) that makes the optimization task harder. For example, in the discrete action space in Fig. (1) (b), moving from defective configurations A to B can be attained via swap moves on a discrete atomistic lattice to navigate via a finite number of paths and reach the global minima at C. On the other hand, for the same task in continuous action space as shown in Fig. (1) (c), there are infinite possible intermediate states and transition pathways possible between any two end states (crystal or configurations) A and B.

In this work, we introduce a scalable RL approach for structure & topology prediction, design, and optimization. This framework termed CASTING (abbreviation for Continuous Action Space Tree search for INverse desiGn) employs a decision tree-based RL algorithm *i.e.* Monte Carlo Tree Search (MCTS)[31, 39, 41]. To navigate efficiently through a high-dimensional search landscape with complex and multiple objectives or rewards, MCTS explores the search space by semi-stochastically sampling (playouts) in the proximity of a node, evaluating and learning its quality in a given search tree. It then takes policy-based decisions to explore the regimes of the search space (i.e., part of a tree) while striking a balance between exploration and exploitation to efficiently reach the target objective *i.e.,* a configuration that maps to our desired material properties. We demonstrate the accuracy, speed of convergence, scalability, and applicability of our CASTING framework across a spectrum of problems (from bulk to low-dimensional, single to multiple components, and search space varying from unit to several large supercells) in the domain of CSP (Crystal structure prediction) and Design. We deploy the CASTING framework to a variety of bulk systems ranging from metal such as Ag to a covalent system such as carbon and explore their stable/metastable polymorphs[38, 46]. We compare the speed of convergence, the accuracy of the best solution, and the sampling quality obtained with the RL approach versus the traditional evolutionary approaches based on genetic algorithm (GA)[9, 13]. We demonstrate the scalability of the CASTING Framework using Ag as a representative system and discuss in detail the impact of the various RL hyperparameters on the structure search. Finally, we employ CASTING for sampling global minima of systems with dimensionalities ranging from 0D to 2D to 3D and across multicomponent systems such as graphene, h-boron nitride, and strongly correlated systems such as neodymium nickel oxide[47].

## Results:

**Crystal Structure Optimization:**

To perform a crystal structure optimization, we represent the configuration or the crystal as either periodic (bulk) or a low dimensional crystal by specifying a set of lattice parameters, basis atoms, and/or atomic compositions of its species. We treat the above-described problem as optimization of the lattice parameters $(a, b, c, \alpha, \beta, \gamma)$, the number of basis atoms ($n$), its positions, and atomic compositions of its species. Thus, any crystal structure is represented as a vector with 6 lattice parameters, and 3 times the number of atom coordinates (x, y, z) with chemical species belonging to each point. MCTS spawns a tree with each node containing a point in the parameter space being searched for and a score indicating the potential to find a promising structure nearby. The root node is initially assigned to random points in the parameter space or seeded with previously known configurations as shown in Fig. 2 (a). To sample a node nearby by perturbing the configurations, we implement different perturbation moves. Mainly 4 types of perturbation (Fig. 2 (c)) moves were used (a) "Add atom" (retaining the composition), (b) "Remove atom" (retaining the composition), (c) "Mutate lattice" (mutation of lattice parameters) and (d) "Mutate atom" (mutation of atomic coordinates). It is to be noted that for the mutation of lattice parameters and coordinates we employ a hypersphere perturbation scheme (see methods section). The radius of the hypersphere is gradually reduced using a gaussian "Depth scaling" function (see methods section & supplementary Fig. S2 (b)). It is also to be noted that the moves that change dimensionality (i.e., size of the system) such as "Add atom" or "Remove atom" are done for only one composition unit. This helps to maintain a parent-child correspondence for a given node. The target objective such as cohesive energies per atom (although any target property computed using MD can be used) of the structures were computed after local atomistic relaxation with the LAMMPS[48] package and the electronic properties such as band-gap were computed using the VASP[49] package.

The optimization with MCTS primarily involves 4 stages starting from a point in parameter space (root Node) and branching out by sampling new parameter sets (crystal configurations) Fig. 2 (a). The first stage involves expanding a node ("Expansion") by sampling new offspring nodes from it by using perturbations (Add atom, Remove atom, Mutate, etc.). Then it is the "Simulation", where the search learns a qualitative score for selected offspring nodes by carrying out random playouts. A playout is basically random exploration near a parent node in the search space by spawning new offspring from it, that are not radically different from the parent but inherits some of its traits instead (see method section). From the overall quality of these offspring's, a measure of a qualitative score of a parent node is obtained. learnings are then backpropagated ("Backpropagation") to the root node for updating the score of the tree. And a "Selection" and further "Expansion" are carried out thereafter. It is to be noted that modified MCTS follows a UCB (Upper Confidence Bounds) (Eq.2) policy for the selection of a node (see method section). The search is conducted till the termination criterion is reached. All the sampled configurations are then mapped according to their stability and potentially good samples are selected based on filtering descriptors[30, 50].

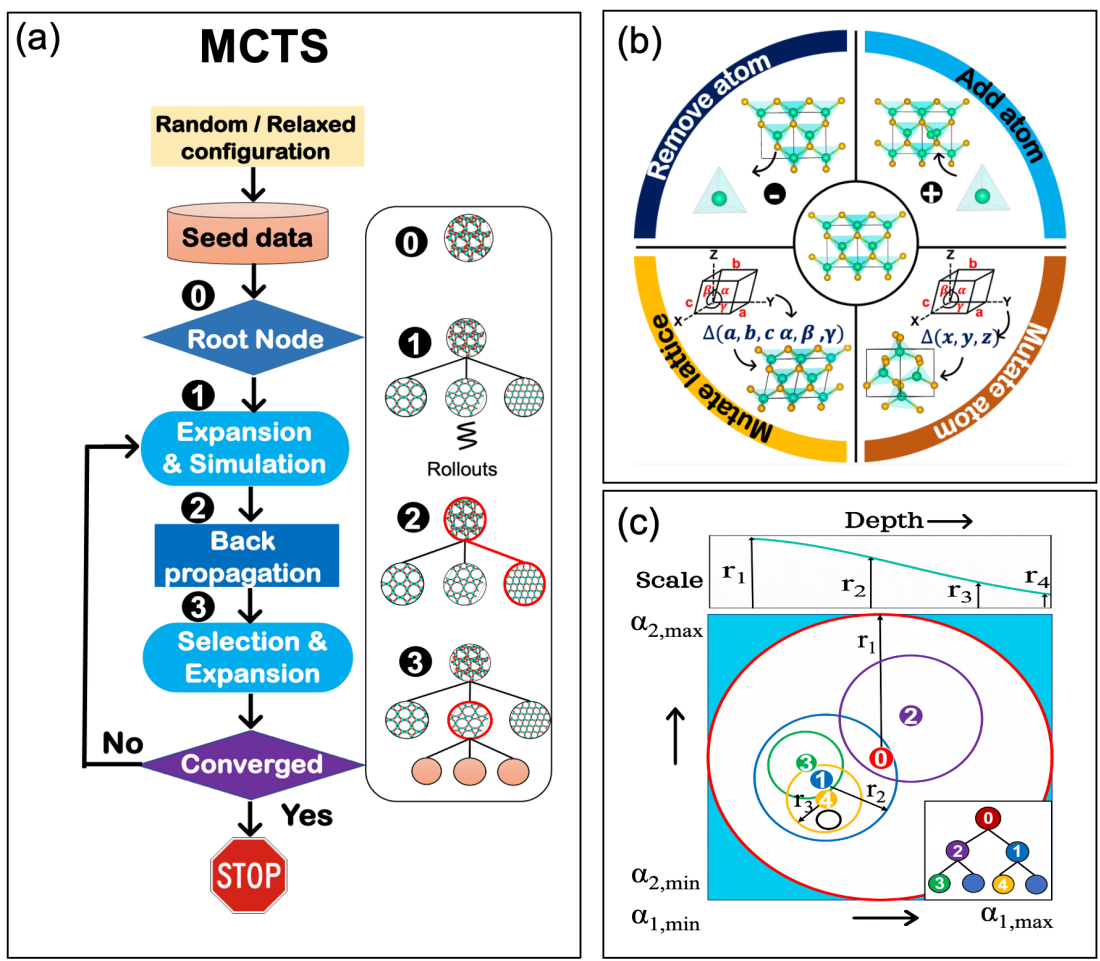

***Fig. 2.*** *MCTS working as crystal structure optimizer. (a) Workflow showing the various stages of MCTS deployed as a crystal structure optimizer constructing a tree search starting from a random or a relaxed configuration as a single node (b) 4 Different types of perturbation moves imparted on a crystal structure*

*in a node as an offspring crystal is created from a parent. (c) "Depth Scaling" scheme, implemented as decreasing radius of hypersphere as the depth of the search tree increases.*

**The CASTING Framework:**

Fig. 3 (a-b) provides an overview of the CASTING framework developed in this work. It has 6 modules that require input from the user. These include (1) The definition of the optimizer (2) selection of target properties to be predicted (3) objective definition or scoring function (4) definition of the crystal system including types of species and number of components (5) simulator or evaluator for the target property (MD or Ab-initio packages) and (6) output options for data analysis and information extraction. An additional "Outputs & Monitor" module provides visualization options for the end user (Fig. 3). The first section requires the user to select the optimizer of choice (RL approach such as MCTS or evolutionary such as GA) and set corresponding hyperparameters that are required with it. In this study, we focus on MCTS as our primary optimizer although we make some limited comparisons to a genetic algorithm-based search in selected cases. The tree hyperparameters that require explicit input from the user are the number of "Head expansion", the number of "Playouts", "Exploration constant", a "Depth Scaling" parameter, and the maximum depth of the tree (see Methods section for details). The selection of target properties that need to be optimized is specified next. The properties can be energetics-based (potential energy, enthalpy, free energy), mechanical (elastic, phonon), electronic (band structure, density of states), and/or thermal to name a few. In this work, we primarily use energy as our target property. Selection of objective function is a crucial step and is entirely dependent on the choice of the optimizer. With MCTS, we use the "UCB" (Eq. 2) as the objective function (see Methods section). The "UCB" itself requires the "exploit" or the "reward" (e.g., configurational energy) to be defined. Additionally, the weights on each "exploit" may be required in the case of multi-objective optimization. Next, the crystal parameters are to be specified. This includes a range for the number of atoms in the simulation cell, lattice bounds range, lattice angle range, chemical species & compositions, and minimum allowed interatomic distance. These parameters define the search space, size, and dimensionality of the optimization. After the target properties, crystal system, and objective function are defined, the user needs to provide corresponding packages for atomistic and electronic calculations (e.g., LAMMPS & VASP package for MD (molecular dynamics) and DFT (Density functional theory) respectively, are used in this study). This part also contains the simulation settings and parameter flags associated with these property evaluation packages. Finally, the "Output options" is for the post-processing section. The user defines the additional outputs such as data formats, visualization monitors, termination criteria, and other metrics that can be used for a quantitative understanding of the quality of a search. There is an additional "Outputs & Monitor" section which provides the user with the flexibility to monitor on the fly, search attributes such as current objective status, tree size, node content, sampled configuration, etc.

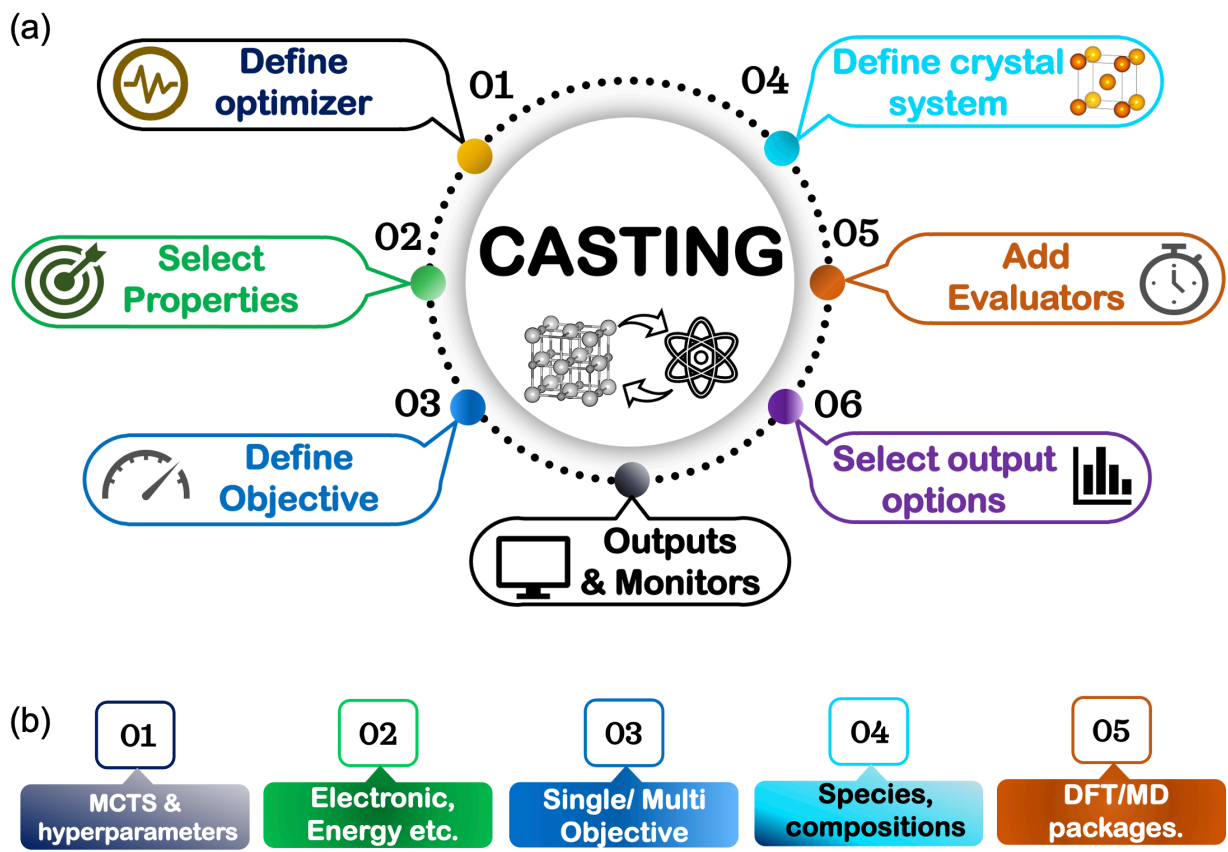

*Fig. 3.* Schematic depicting the workflow of the CASTING framework for performing inverse design – *(a) User interface for specifying various IO settings leading to a different set of operations at the front-end of CASTING. These include (01) Defining the type of optimizer (02) Selection of properties to be predicted (03) Objective definition (04) Definition of the crystal system or configuration, (05) Evaluators (MD or Ab-initio packages) for computing the rewards or score, and (06) Output options. An additional "Outputs & Monitor" module is available for visualization. (b) Additional input options associated with each of the operations specified at the front end in (a) – this includes (01) MCTS search and associated hyperparameters (02) target properties to be computed (03) single or multiple objectives (04) single or multicomponent or types of species (05) classical MD or electronic structure simulator to evaluate properties.*

**Applications of CASTING:**

We have applied our CASTING framework to a set of very relevant and challenging problems in the domain of CSP (Crystal structure prediction) and design. The emphasis is on the demonstration of accuracy, speed of convergence, scalability of the workflow across a diverse range of problems at different dimensionalities. To test the scalability, and speed of convergence, we first use an example of a metal such as Ag with fewer polymorphs and a smaller number of known local minima in their energy landscape. We then extend our approach to predict covalent system such as Carbon that exhibits a diverse range of metastable states and polymorphs and compare the RL-based search with a more traditional evolutionary approach. The aforementioned problems all deal with a bulk (periodic) system. We next move beyond bulk (periodic systems) to explore dimensionality effects on our workflow. Primarily, we explore two different classes of systems- a 0 D (cluster) single component system such as gold (Au) for representative sizes and global minima for 2D binary systems such as C-H (Graphane), and Boron Nitride (h-BN). Finally, to explicitly explore the compositional variance induced metastability, we deploy CASTING to explore the compositional space of doped Neodymium Nickel Oxide (NNO) and their impact on a representative electronic property such as bandgap. The hyperparameters used for searches involving each of the above material systems are specified in **Table 1**.

*Table. 1* *CASTING hyperparameters used for different material systems.*

| Task | Maximum depth | Head expansion | Exploration constant |
|---|---|---|---|
| Metal polymorphs (Ag) | 6, 12 | 5,10,100,1000 | 0.14,1.4,14 |
| Metastable Polymorphs of Carbon (C) | 12 | 10 | 4 |
| Gold (Au) nanoclusters (13, 20, 40 atoms) | 12 | 10 | 1 |
| Boron Nitride h-BN (2D) | 12 | 10 | 2.4 |
| Graphane (CH) (2D) | 12 | 10 | 1.4 |
| H Doped -Neodymium Nickel oxide (NNO) | -- | -- | |

1. **Exploring the scalability of CASTING framework using an example of metal polymorphs:**

Silver (Ag) is a well-studied metal and is known to have only a few metastable polymorphs (e.g., hcp, stacking faults, etc.) with the fcc as the most stable or ground state in its bulk form. We utilize Ag as a representative test case to evaluate the scalability of our framework. Any structural search performed with a decision tree such as MCTS primarily depends on the two aspects of the search parameters. (a) specifications of the crystal parameters (size, lattice parameters), and (b) hyperparameters that control the construction of the tree.

We first explore the impact of the crystal input parameters on the performance of our RL approach. Given that the solution is known (*i.e.*, the lattice petameters and atomic coordinates of ground state fcc structure), we set the search bounds of the lattice parameter in terms of percentage deviation ($\delta$) from its stable counterpart. For example, an increment in the bounds by 30% means a lattice vector range of [0.7***a**,1.3***a**], where **a** is the lattice vector of the pure fcc for a given size of supercell. It is also intuitive that an increment in lattice bounds means an increase in the search space area as well - this not only introduces degeneracy in the obtained solution but may affect the overall search quality as well as shown in Fig. 4(b). We first start with a 4-atom search to test the typical convergence profile of the MCTS optimizer and compare it with a purely random search with local minimizations of the configurations to get an idea of the qualitative threshold (Fig. 4 (a)). We use an EAM type empirical potential [46] and set the lattice parameters bounds deviation($\delta$) to be 30% (See Table. 1 for the hyperparameters). A LAMMPS[48] simulation package was used for the evaluation of the structural property (energy). We find that allowing atoms to approach closer during the search (*i.e.*, specifying a lower value for minimum inter-atomic distance criteria) allows the RL to more exhaustively explore the search space - from high energy regimes and overcome energy barriers) and helps in overall convergence.

Fig. 4(a) shows that our MCTS search reaches the optimal solution in fewer evaluations compared to the random sampling – the solution quality with MCTS is also better *i.e.* lower in configurational energy. The stacking of the final predicted structure corresponds to an fcc fingerprint. The energy difference of the final solution from MCTS to that of the pure fcc is negligible (<<1meV). Since we are growing a tree of finite size while exploring search space, it is expected that a significant change in the search space size (area) might affect the performance of the search. We define a search area to be the magnitude of vector cross product between the upper and lower bound of the lattice parameters vectors. To test this dependence, we spawn 3 trees using the same root node with different head expansions (h) and depth (d) (Fig. 4 (b)). For a tree with less width (head expansion) (h=5,d=12), with the increase in the search area, the performance drops rapidly since the size of the tree is not adequate to cover the entire search space. As the width of the tree increases (h=10, d=12) the performance becomes much better for lower value areas of the search space. However, we do see a general decline in the performance, with an increase in the search space area. This is because, in a continuous actions space, an increment in the search space area introduces innumerable configurational possibilities in the energy landscape. Thus, a greater number of iterations are required to explore it. At the same time, it is also obvious that a shallow tree (less depth) (h=10, d=6) also results in poor performance. As the tree depth increases, the search mostly exploits branches with promising nodes in the tree. A shallow tree restricts the search from exploitation, resulting in delayed or no convergence at all.

We next test the scalability of the CASTING workflow by testing the convergence speed and the energy per atom difference for convergence towards a unit cell of fcc (4 atoms), a supercell of 2*2*2(32 atoms), a supercell of 3*3*3 (108 atoms), and a supercell of 4*4*4 (256 atoms). The width and the depth of the search tree are kept fixed (h=10, d=12). We also select a wide range of the search bounds deviation ($\delta$) from 10 to 30% deviation for testing. We perform 6 independent trial searches (initializing the root node of the tree at different points in search space) for each of the cases with the maximum number of iterations kept at 20000. For the best solution from each of these trials, the distribution of energy difference from its fcc supercell counterpart, and also the corresponding difference of the structure in terms of lattice parameters and stacking have been shown in Figs. 4 (d-e). To determine the similarity of the atoms to that of an fcc stacked lattice we used bond order-based parameters based descriptor ($Q_2$, $Q_4$, $Q_6$)[51] and coordination number (CN). It can be observed that for each of these sizes there is an optimal bounds deviation ($\delta$), for which the search gives the best performance (less variation in final energies and very close to the target) (Fig. 4(d)). It is also to be noted that as we move higher either in size of the system or the bounds deviation ($\delta$), there is a tendency to achieve solutions that have vastly different lattices from the orthogonal supercell, but atoms are stacked in an fcc motif (Fig. 4(e)) with energies extremely close to the target solution. The effect is more prominent with changes in bounds deviation ($\delta$). These primarily are two contributing factors for MCTS obtaining these degenerate solutions, (1) With an increase either in size or dimension(size) of bounds deviation ($\delta$), the search constraints get lighter allowing atoms to arrange themselves in fcc motif while not having an orthogonal lattice (2) With an increase in the bounds, the corresponding area of the search space also increases, which allows MCTS to explore higher energy regimes of the search space (see supplementary information Fig. S1(c)) causing it to find these energetically close degenerate solution while severely delaying the final stages of the convergence (reaching to the exact orthogonal structure). There is also a dependency on the size of the tree as discussed earlier. For example, with 4 atoms at $\delta = 10\%$, the atom can only arrange themselves in an orthogonal fcc unicell, thus the best solution is obtained. With $\delta = 20\%$, the atoms do not have the flexibility of getting degenerate solutions, and also the size of the tree relatively is large for a given search space area. Hence the search could not get to solutions within fixed iterations (20000) and the energy distribution is wide (Fig. 4(d)). For $\delta = 30\%$, the degeneracy can be seen, thus the energy distribution becomes much better owing to these solutions. Similar nonmonotonicity in performance can be observed for the other sizes too. The overall performance, for the given size of the tree (h=10, d=12) is optimum at $\delta = 30\%$, for all the dimensionalities (system sizes). Note that with the increased dimensionality (Fig. 4 (d)), the best solution obtained by MCTS for each case has a range of energy difference < 0.15 meV, indicating the ability of the MCTS optimizer to scale to the dimensionality as high as 774 (256 atoms * 3 cartesian coordinates + 6 lattice parameters) while maintaining a considerable solution accuracy. While for a random search the performance deteriorates considerably (see supplementary information Fig. S1 (b)).

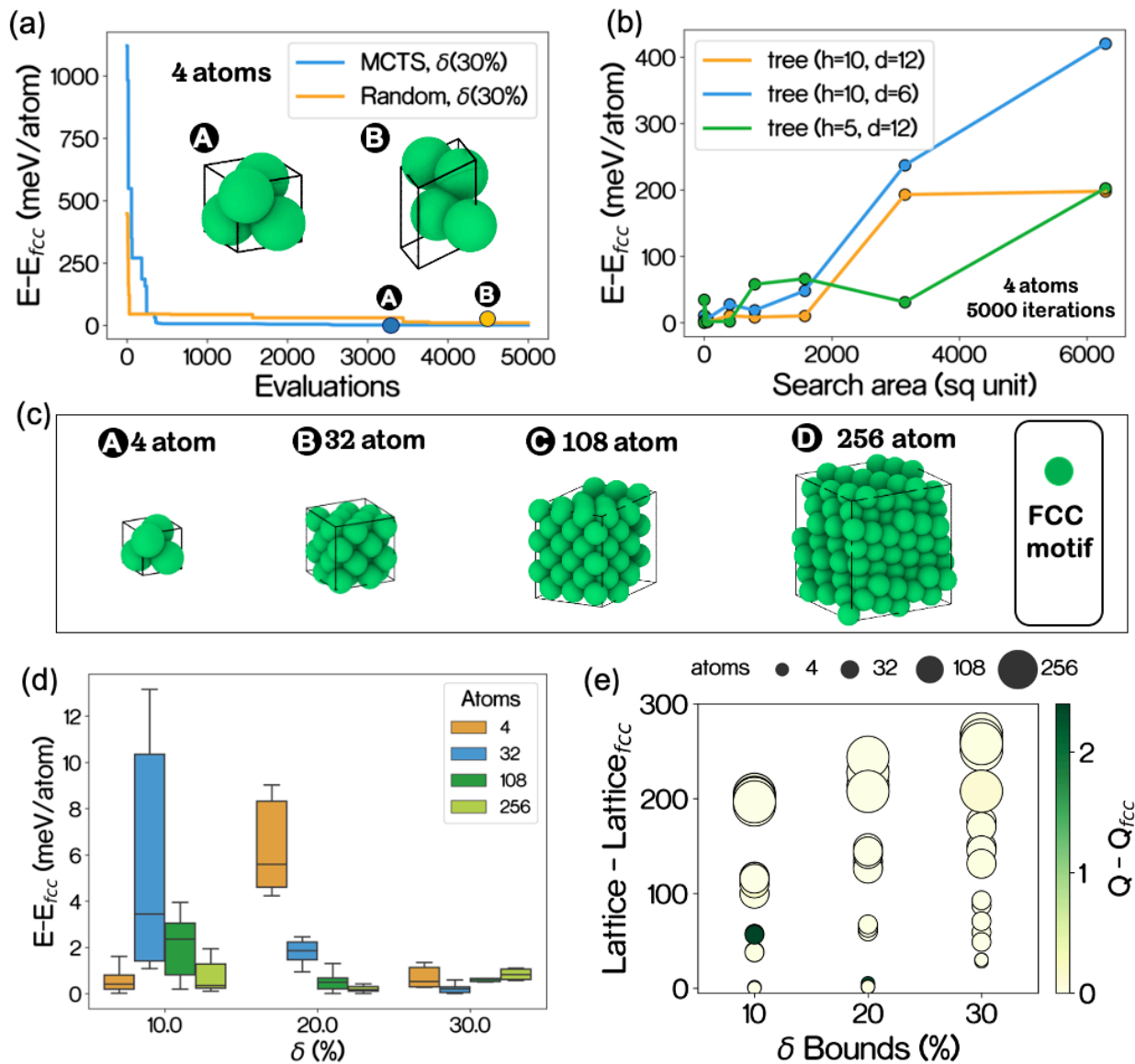

*Fig. 4.* Exploring the performance and scalability of CASTING framework using an example of metal polymorphs (a) Comparison of the speed of convergence and difference in energy from the best available

solution (Ag$_{fcc}$) between random sampling and MCTS optimizer for 4 atom system of Ag. (b) Performance of the MCTS optimizer (for different sizes of tree) for the problem in (a) as the area of the search space changes. (c) Effect of dimensionality on the predicted crystal structure for different system sizes. (d) distribution energy difference (from fcc) (meV/atom) of the best solution obtained (in 20000 iterations) for 6 independent trials on different sizes of the system with increasing lattice parameter bounds ($\delta$) from a relaxed orthogonal supercell Ag (fcc). (e) Structural variation for the different minima obtained from the independent trials (as in (d)) in terms of changes in lattice parameters (from a relaxed orthogonal fcc supercell) and atomic stacking (difference from a pure fcc) for different sizes and lattice parameter bounds($\delta$).

Next, we explicitly explore the different tree hyperparameters and analyze their effect on the convergence and overall sampling quality as shown in Fig. 5. The maximum number of iterations was kept at 2000 and the starting point (root node) of the search was kept the same for all cases. The # of atoms range was fixed at 4 atoms and a bounds deviation ($\delta$) of 30% was maintained. In Fig. 5 (a), we show the effect of the increasing head expansions for the tree construction on the overall sampling and convergence of the search. The head expansion of the MCTS is somewhat comparable to generating an initial population in the evolutionary approaches. To start with, one would want to have minimal sampled points that cover the search space uniformly. Further branching out from those points helps the search to converge faster. Too many head expansions will generate redundant points in the same regions of search space causing the MCTS to explore unnecessarily more before reaching a converged solution resulting in an energy distribution with a high mean (Fig. 5(a)) and a typical slower convergence. The converse is true for a very smaller number of head expansions which might cause the search to get stuck in a certain region of the search space and may completely obstruct its convergence. We next look at the effect of playouts (Fig. 5 (b)). Playouts are basically random perturbations on a node to get a quantitative idea of how much likely a node is to yield a good offspring upon further exploration. From the perspective of sampling, it is evident that there is an optimum for the number of playouts required. Too much of a playout will unnecessarily increase the number of iterations thus resulting in a slower convergence and too less of a playout might result in incomplete knowledge regarding any given node leading it to converge at a slower pace as well.

The exploration constant is another crucial parameter for the UCB (see methods section - Eq.2) setting as well as an important parameter that controls the exploration of the tree. For too small of an exploration constant, the tree will greedily pick the nodes with good objective value only making the search confined to a certain region of the search space (Greedy Search). This can have an adverse effect on overall convergence. On, the other hand, selecting a too large constant will make the search to be effectively random. So, a proper selection of exploration constants can help the search to converge efficiently in relatively few numbers of expensive objective function evaluations (Fig. 5(c)). The final hyperparameter that we explored the effect of is the "depth scaling". For any MCTS search, as the depth of the tree increases the parameters at the nodes are expected to be closer to the converged solution than that of a node residing at a higher depth. This is also indicating that the search is moving towards an exploitative phase and thus a scaling of the sampling window is necessary. Otherwise, it might deviate the search from moving towards convergence. We use a gaussian type depth scaling scheme (see Methods section, supplementary information Fig. S2). From Fig. 5 (d), we see that there is a slightly slower convergence for both higher and low values of "a". A low value of "a" causes the search to become too much exploitative at a shallow depth of the tree. Since it samples only degenerate solutions in a small region of the search space while a high value of "a" prevents it from being exploitive in a tree with high depth when it is required to do so.

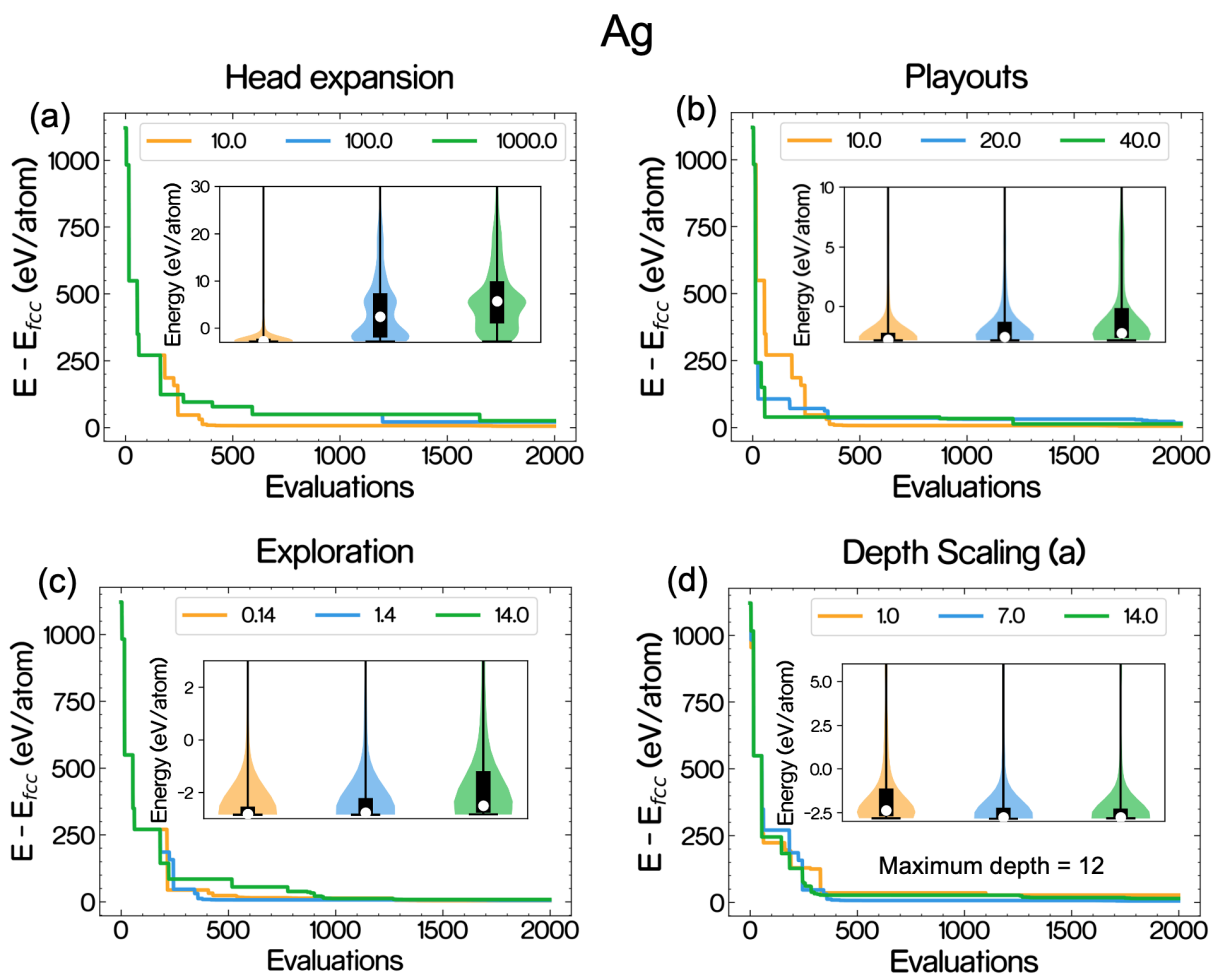

*Fig. 5. Effect of tree hyperparameter on the sampling, convergence, and solution quality of Ag polymorphs. (a) Shows the convergence and energy distributions for different head expansions. (b) Shows the convergence and energy distributions for different playouts used. (c) Shows the convergence and energy distributions for different head expansions used. (c) Shows the convergence and energy distributions for different head exploration constants used. (d) Shows the convergence and energy distributions for different depth scaling factors "a" used.*

## 2. Exploring the Diverse Metastable States and Polymorphs of Carbon using CASTING:

We next explore another system which has a high degree of metastability *i.e.,* has a large number of local minima in its energy surface. Carbon is known to have a diverse range of allotropes, in terms of size, property, and structural diversity. This makes it a suitable test system for benchmarking the sampling quality, accuracy, and speed of convergence of the CASTING framework. Since it is already known that graphite and diamond (at high pressure) are the two most stable allotropes, we set them as our target solution. We start with 3 different search cases (a) CASTING (b) genetic algorithm (GA)[9] (c) random search with local minimization of the structures - the atom# is in the range [2,10], lattice vector range [2 Å, 8 Å], and lattice angle range [$60^0$, $120^0$]. The Tree hyperparameter settings are given in Table 1. The empirical LCBOP[52] potential along with the LAMMPS simulation package for local minimization of the configurations and calculation of energy.

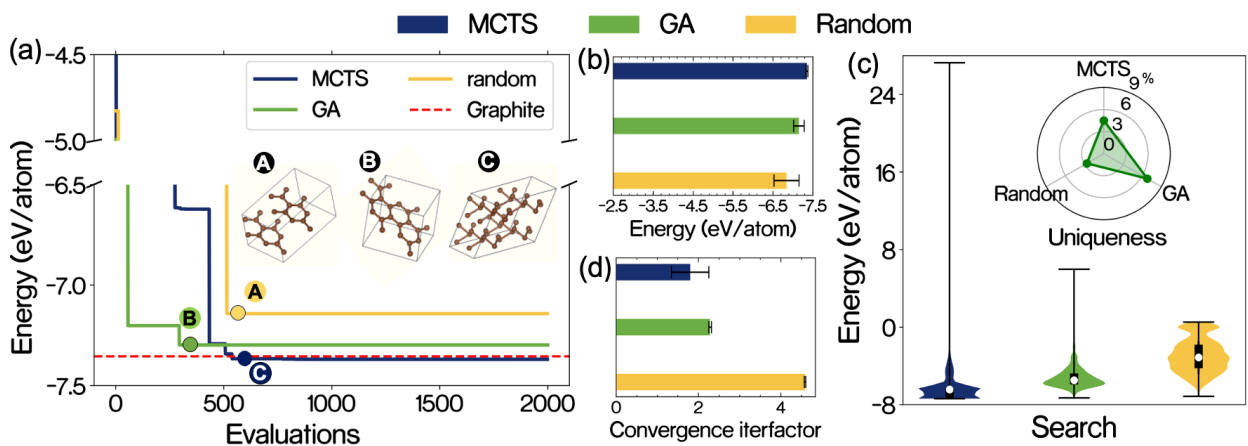

*Fig. 6. Comparison of structure prediction for carbon polymorphs with an empirical potential model[52]. (a) Best Convergence of MCTS, GA, and random sampling out of 4 independent trials. (b) Mean the best*

*solution obtained for MCTS, GA, and random sampling (c) Typical energy distribution of the sampled configuration during an independent run for MCTS and GA optimizer and their overall uniqueness (d) Average iteration factor for convergence for different optimizer algorithms used.*

From the results of 3 independent trials (Fig. 6 (b), (d)) and the best solution for each case (Fig. 6 (a)), it is very clear that the MCTS optimizer in the CASTING framework not only converges faster to the solution Fig. 6 (d) (The 'convergence iterfactor' is the normalized number of iterations taken for the convergence of the search), but the quality (the energy per atom) is also better (Fig. 6 (b)). We also compare the property (energy per atom) distribution of the configurations sampled using MCTS and GA optimizers (Fig. 6 (c)). Clearly, MCTS tends to sample more configurations in the lower energy range as compared to GA, but the overall uniqueness of the sampled configurations is less as compared to the GA (Fig. 6 (c)). This is indicative of the fact the MCTS tends to sample more similar polymorphs near the global minima to reach the absolute best solution (exploitive) since most of the PES of empirical[52] potentials have degenerate solution of the same structure (Graphite in our case) with a very minute difference in energy. Which sometimes hinders more exploratory type search algorithms such as GA to reach the absolute solution. On the other hand, the GA has a slight upper hand in terms of sampling more diverse polymorphs because of its exploratory nature.

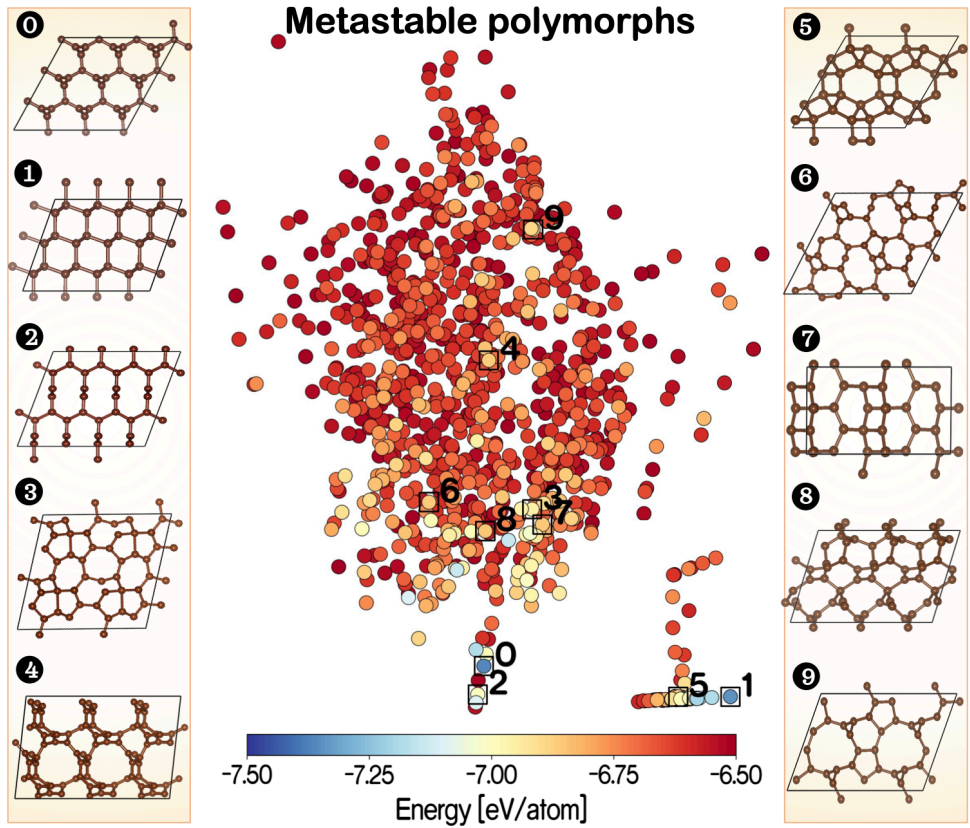

***Fig. 7.*** *ISOMAP representation of Order Parameters ($Q_2$, $Q_4$, $Q_6$) & Coordination Number) Based feature vectors of bulk metastable polymorphs of "C" sampled using CASTING framework with LCBOP[52] interatomic potential across a range of external stress spanning from 0-120GPa.*

It is to be noted that the MCTS can also be made exploratory in nature by incrementing the exploration constant "C" in the UCB equation (Eq.2) (see methods). By implementing the same for the Carbon polymorphs, we search with our CASTING framework for metastable phases of carbon polymorphs at different external pressure ranging from 0 to 120 GPa. To find out the unique ones amongst the multiple different structures sampled with MCTS, we adopt a two-step method. Our solution contained a lot of variants of graphite polymorphs. Therefore, we first apply a graph neural network-based characterization workflow[30] to isolate the 2D layered polymorphs from bulk structures. Next, we filter out the unique ones from the bulk configurations using order parameters ($Q_2$, $Q_4$, $Q_6$)[51] + coordination number feature representation of the bulk configurations and an unsupervised agglomerative clustering[53] technique (see supplementary section S. 1). From the ISOMAP representation[54] feature vectors of the unique bulk polymorphs (Fig. 7), the MCTS optimizer not only sampled a large number of (~1.2K) diverse metastable polymorphs of carbon but also across a wide energy window (~ 1eV). It is also to be noted that MCTS managed to sample the diamond structure (Fig. 7 configuration 1) that exists at higher energy value as compared to the global minima graphite. In the phase diagram of carbon[55], the graphite polymorph is stable at regular thermodynamic conditions whereas the diamond polymorph exists under extreme pressures, which, makes the diamond polymorph metastable at regular thermodynamic conditions. Since there are exponentially many local minima introduced as the overall energy window of the search increases[1], thus discovering diamond becomes difficult.

## 3. Beyond bulk or periodic systems – Exploring dimensionality effects on CASTING's search performance:

Low dimensional materials with their high surface to volume ratios present a unique opportunity to tap into properties that cannot be attained in the bulk form[18, 56]. As the dimensionality of the atomic particles enters the regime of non-periodicity, the additional abundant surface (nanoclusters, layered materials), weak van der Waals interaction between the layers (2D) leads to electronic changes[57], that begins to play a dominant role in displaying exotic electronic and optical properties having potential in a multitude of applications such as semiconductor electronics[56, 58, 59, 60], transport[61], biotechnology[62], medicinal applications[63], systems with mechanical responses[64], etc. Like any atomistic system, the complexity of predicting low dimensional metastable phases increases with the size of the system. For zero-dimensional clusters, the variability in atomic packing[25], a wide energy landscape (from gaseous phase to condensed), and presence of isomorphism makes it challenging to efficiently explore the search space. While for layered 2d, materials, a weak Vander Walls interaction, variability in coordination environment makes it possible to form numerous local minima that are hard to explore but with potential of having desired properties of interest. It is also worth mentioning that the existing knowledge of the nanoparticle structures does not reach the atom-level resolution from experiments[65]. Traditional diffraction techniques are more suitable for periodic crystalline structures. Thus, for decent accuracy of prediction, comparison of data from multifarious techniques is required[65]. In this regard, CASING can provide a unified platform for predicting global and local minima of these low dimensional systems to bridge this gap between theory and experiments.

## 4. 0D – Exploring the size dependent diversity in Gold (Au) nanoclusters:

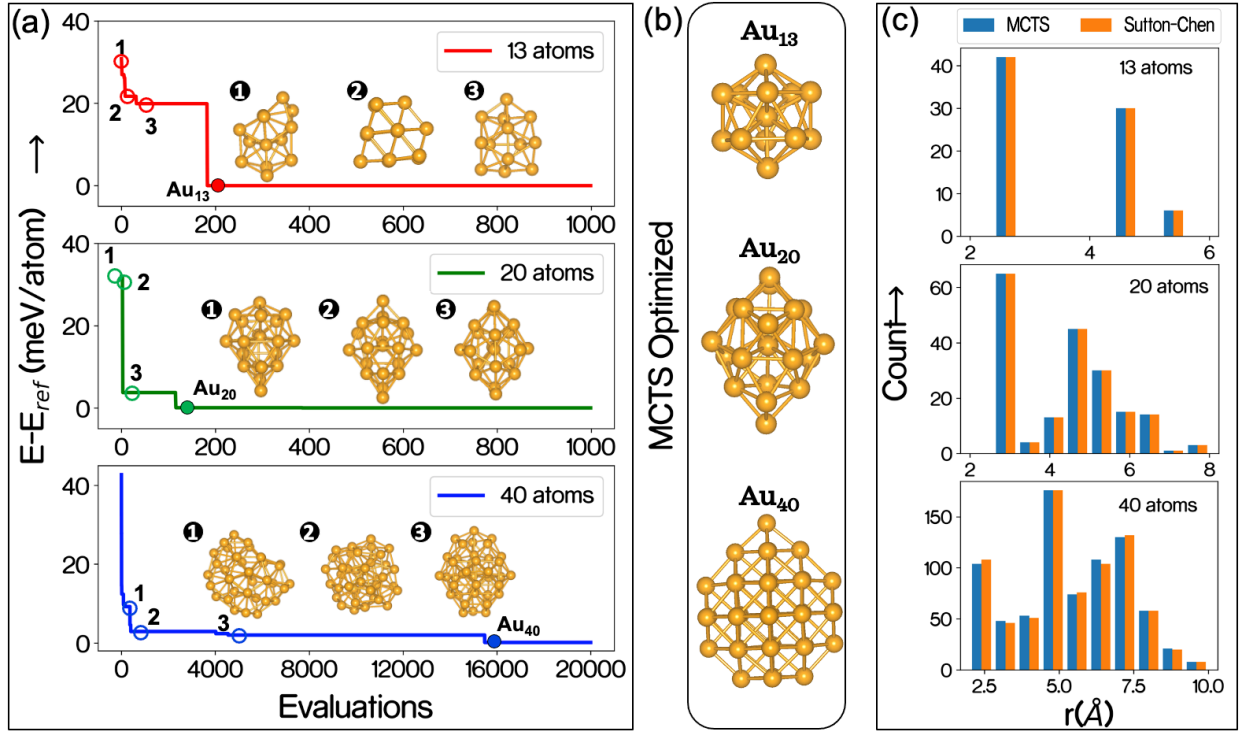

*Fig. 8. (a) The convergence of MCTS optimizer for the sampling of gold (Au) nanoclusters of different sizes (13,20, and 40 atoms) (b) shows the global minima obtained by MCTS for each case (c) Comparison of the global minima obtained by MCTS to that of known Sutton-Chen global minima[66] in terms of pairwise distance between the atoms for 13, 20, and 40 atoms respectively.*

We start by employing CASTING for the search of global minima of gold (Au) nanoclusters. Au nanoclusters due to their versatile applicability, have drawn significant attention over the years[60, 67, 68, 69]. Computationally, most of the global minima of these nanoclusters of different sizes have been extensively explored[66]. Yet due to their relevance in modern-day material science[60], the optimization of these nanoclusters is of great interest. In this work, we use Sutton-Chen (10-8)[70] interatomic potential to recover the known global minima of gold (Au) nanoclusters[66] having 13 Atoms, 20 atoms, and 40 atoms respectively. Fig. 8 (a) shows the convergence of the MCTS optimizer for the 3 representative sizes of clusters used in this study. The tree hyperparameters for all the cases of search (see Table. 1) kept being the same. A LAMMPS package was used for the local minimization of the atomic configurations. For the 13-atom cluster which is known to have an icosahedron structure as the global minima, the MCTS optimizer takes ~150 evaluations to converge to the solution which is considerably less. As their dimensionality increases with an increase in size, the iteration taken by the MCTS optimizer to converge to the global optimum also increases expectedly. For the 20-atom size is around ~300 iterations and that of 40 atoms for which to reach the global minima, MCTS takes ~20000 evaluations to reach the optimum solution. We also compare the global minima obtained by MCTS (Fig. 8(b)) with the known global minimas[66] for each of the sizes in terms of their structural features (pairwise interatomic distances Fig. 8(c)). From Fig. 8(c) it can clearly be seen that, apart from being similar in terms of energy, MCTS optimized structures obtain identical structural similarity to their know counterparts. The overall results display the efficacy of the CASTING workflow in successfully scaling down from the bulk system to 0D systems while acknowledging the fact that the cluster systems are more difficult to optimize than their bulk counterpart because of additional degrees of freedom.

## 2D – Exploring the global minima of two-dimensional Boron Nitride (h-BN) and Graphane (CH):

We next test the performance of the CASTING workflow in sampling two-dimensional (monolayer) systems with a richer compositional degree of freedom such as hexagonal Boron Nitride (h-BN) and Graphane (C-H). h-BN is an exceptional insulator with a direct wide bandgap of ~ of 5-6 eV[71, 72, 73]. Being insulating and transparent, it has the potential of becoming an exceptional substrate for the synthesis of Graphene[73, 74], thus being also referred to as "white graphene"[71]. The h-BN has covalently bonded Boron (B) and Nitrogen (N) atoms that crystalizes in a hexagonal $P6_3/mmc$ space group. On the other hand, Graphane is the hydrogenated version of conductive semi-metal Graphene[75]. It is a fully saturated $sp^3$ hydrocarbon with a 1:1 stoichiometry of C:H. Unlike Graphene, Graphane lacks the Dirac cone and also has an indirect bandgap of ~5.4 eV[76], hence behaving like an insulator. Still, its discovery, and structural attributes[77] have unraveled new paradigms in design of new semiconducting counterparts[76] with exotic electronic properties. From the structural perspective, Graphane has two known conformations are of Chair and Boat type[77] with Chair (P3m1) being the global minima. For both systems, the knowledge of the various intermediate and key transition states can not only unravel crucial aspects of metastability but may also prove insights into the synthesis of these and other similar systems[76] – we deploy CASTING to explore the search for the monolayer h-BN and the chair polymorph of Graphane.

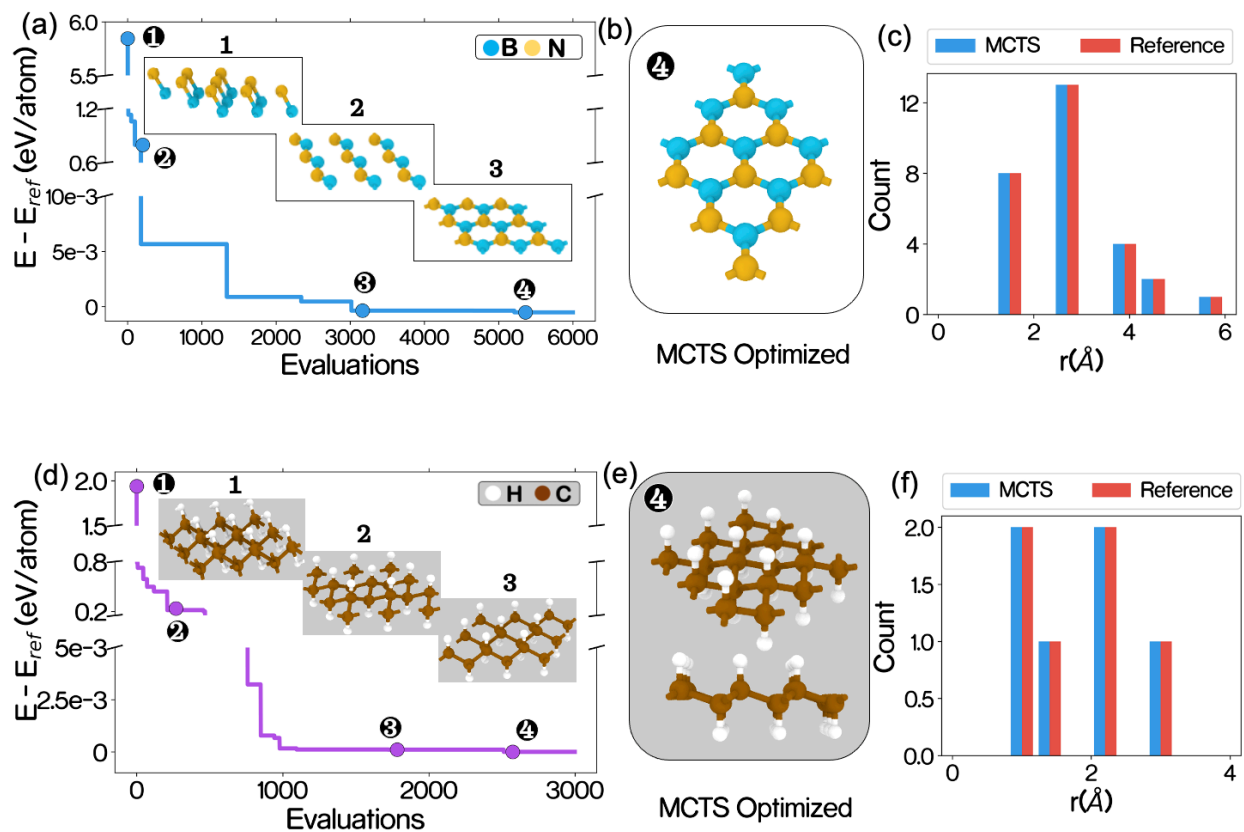

***Fig. 9.*** *Exploring 2-d polymorphs with CASTING (a) The convergence of MCTS optimizer for the sampling of Hexagonal Boron Nitride (h-BN) (b) shows the global minima of h-BN obtained by MCTS (c) Comparison of the global minima obtained by MCTS to that of known global minima of h-BN in terms of pairwise distance between the atoms. (d) The convergence of MCTS optimizer for the sampling of Graphane (CH) (e) shows the global minima of Graphane obtained by MCTS (f) Comparison of the global minima of Graphane obtained by MCTS to that of known global minima in terms of pairwise distance between the atoms.*

To describe the interactions between B & N atoms we use an extended Tersoff [73] type potential while for C & H, the popular AIREBO potential[78] is used. We search for the unit cell of both h-BN and chair polymorph of Graphane. We conduct our search for the same number of atoms in the unit cell and keep the deviation in the lattice parameter ($\delta$) from the known global minima to be 20%. Fig. 9 (a) (d) shows the convergence of the MCTS optimizer with the number of relaxation evaluations of LAMMPS for h-BN and Graphane respectively. The total number of evaluations taken by the MCTS optimizer to converge to the global minima for h-BN is around ~3K while that for Graphane is ~2.5K. The target solution reaches the exact energy per atom value compared to its reference counterpart for both cases (Fig.

9 (a) (d)). We also compare the global minima of h-BN obtained by MCTS (Fig. 9 (c)) to the known global minima[79] in terms of pairwise interatomic distances. The excellent match indicates that CASTING has achieved perfect accuracy structurally. A similar observation is seen in the case of the Graphane search (Fig. 9 (e), (f)). In addition to having more degrees of freedom, the compositional space adds more challenges to the optimization problem due to the inclusion of multiple species. Note that the increased diversity in the 2d conformations and a richer polymorphism also make it harder to reach the global minima for any search algorithm. This presence of multiple local minima in low dimensional systems translates into a higher number of iterations for MCTS to converge to the solution when compared to the bulk systems.

**Exploring the Compositional Space of Doped Neodymium Nickelate (NNO) using CASTING – Elucidating the Correlation between Metastability and Resistance States:**

We next deploy CASTING to explore an even more complex compositional landscape of a multi-component system *i.e.* perovskite nickelates doped with hydrogen and elucidate the relationship between metastability in doped NNO and their resistance states. Perovskite nickelates systems such as Neodymium Nickel oxide (NNO) can exhibit electronic properties that have immense potential in a multitude of applications [47, 80]. The ground state NNO ($NdNiO_3$) is an orthorhombic perovskite structure with Ni atom bonded to O atom forming a corner-sharing $NiO_6$ octahedra[79]. A strongly correlated system NNO, however a metal at room temperature (see supplementary Fig. S3 (a)), the addition of electron donors (H) in the lattice changes electrical conductivity extensively[47]. This makes it an exceptional candidate for being applicable in brained inspired computing[47, 81]. Additional donated protons from H interstitials to the Ni not only impact its resistivity severely but also induces a complex potential energy surface with a plethora of local minima (metastable states). Additionally, there are two inequivalent O sites in the NNO lattice[79] providing permutational variability towards the location of H atoms. This makes it hard to locate the optimal position of the hydrogen (dopant) atoms in the lattice in search of favorable metastability for resistive switching. The task tends to become more challenging with an increasing concentration of dopants as the number of possible metastable states tends to grow exponentially.

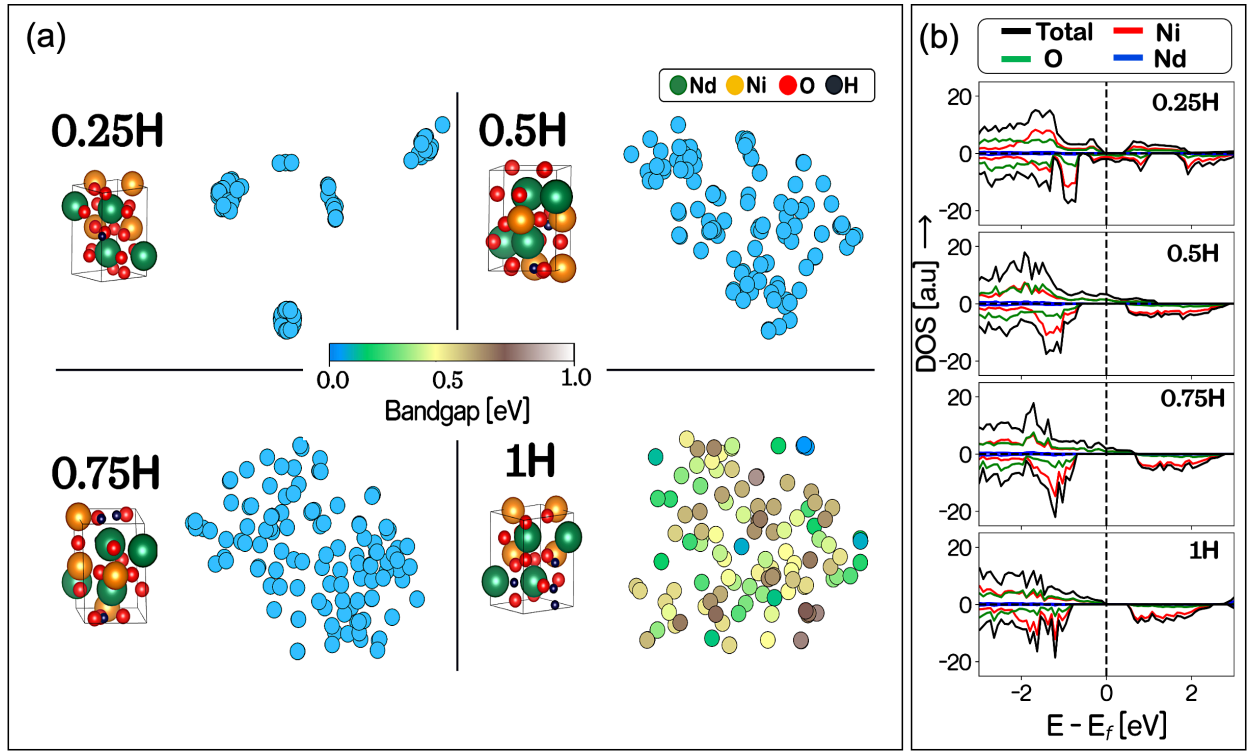

*Fig. 10.* *Exploration of the configurational space of hydrogen doped Neodymium Nickel Oxide (NNO) with CASTING framework. (a) Shows the t-SNE (t distributed stochastic neighbor embedding) plot for SOAP feature representation of the sampled metastable polymorph at different concentration of hydrogen doping and their corresponding band gap magnitudes. (b) the typical density of states of sampled configurations at doping concentrations of 0.25H, 0.5H, and 0.75H, respectively.*

To begin with, we select 4 concentrations of hydrogen doping 0.25H, 0.5H, 0.75H, and 1H respectively (Fig. 10 (a)). We assume that there will be distortions in the NNO lattice upon insertion of H in it, the symmetry of the fundamental NNO lattice does not get broken even after ionic relaxation in VASP. So, during the sampling, we do not apply any external perturbation to the NNO lattice instead we move the H atoms through the lattice by perturbing its location. This allows us also to find possible locations or H sites in the lattice that alters the electronic structure by creating new eigenstates (Fig. 10 (b)). A VASP package was used for structure relaxation and electronic calculation (see supplementary information S. 2 for details). It is intuitive that with the increase in the concentration of doping the possibility of having unique metastable states increase drastically. This can also be observed in Fig. 10 (a). From the t-SNE (t distributed stochastic neighbor embedding) plot of SOAP[50] feature vector representation of the structures having a doping concentration of 0.25H (Fig. 10 (a)), the distinction of the polymorphs in the feature space is not very conspicuous. As the doping concentration increase, the number of discrete and diverse polymorphs tends to grow. It is also very interesting, that the polymorphs having a doping concentration less than 1H, tend to show similar metallic behavior. As the doping concentration reaches 1H, the energy eigenstates vanish near Fermi energy (Fig. 10 (b)) indicating a semiconducting behavior of the polymorphs. The trend persists for almost all the polymorphs sampled at this concentration. This application demonstrates the flexibility of our CASTING towards accurately performing tasks that go beyond simple crystal structure prediction while targeting specific properties of interest in complex material science problems.

## Discussion:

In summary, we introduce CASTING which is a workflow that implements a continuous action space tree-based RL search algorithm for crystal structure prediction (CSP) in a high-dimensional search space. We discuss the important algorithmic modifications that are needed in the MCTS to successfully apply it to continuous search space inverse problems associated with structure and topology predictions. To showcase the efficacy of the CASTING framework, we apply CASTING to a wide range of representative systems – single-component metallic systems such as Ag and Au, covalent systems such as C, binary systems such as h-BN and C-H, and multicomponent perovskite systems such as doped NNO. We demonstrate the scalability, accuracy of sampling, and speed of convergence of CASTING on complex material science problems. We discuss the impact of the various RL hyperparameters on search performance. CASTING is also deployed to sample stable and metastable polymorphs across systems with dimensionality ranging from 3-d (bulk) to low dimensional systems such as 0-d (clusters) and 2-d (sheets). Comparisons to other metaheuristic search algorithms such as genetic algorithms and random sampling are also shown – the MCTS is demonstrated to have a superior performance in terms of the solution quality and the speed of convergence. We expect MCTS to perform well, especially for complex search landscape with multiple objectives, multiple species, and multi-dimensional systems. Overall, we successfully demonstrate the development and application of powerful RL techniques such as MCTS for inverse materials design and discovery problems related to structure and topology predictions.

## Methodology:

**Monte Carlo Tree Search (MCTS) in continuous Action Space:**

Traditional vanilla Monte Carlo Tree Search (MCTS) has been applied to many materials' science problems[32, 82, 83] involving discreet spaces. But the continuous actions space adaptation for the crystal structure prediction requires additional modifications. We have introduced the following to the MCTS to enable its application for continuous search space problems. These include:

**Enhanced Exploration and Degeneracy Protection:** When performing a search of a very large phase space there can be a multitude of problems that arise which if not accounted for will result in the optimizer spending iterations on unnecessary solutions. In the case of crystal structure searches, there are two problems with degenerate results that can arise. First, the optimizer can have two branches that initially start at two different positions in the phase space, yet they will converge into the same location. This is effectively the algorithm retracing its steps repeatedly. The second problem which is more common in structural searches is that the natural entropy of the atomic positions can create many degenerate minima. For example, if one takes all the atoms in a structure and simply translates it a few angstroms in one direction the energy of the system has not changed (Translational invariance). As a result, when performing these searches, one may find a different parameter combination that results in an identical crystal structure. This degeneracy translates into MCTS spending computational cycles on solutions it has already seen before. We define a uniqueness function on the exploration side of the node selection rule to avoid degeneracies in the search space. For situations where we simply wish to limit two branches from approaching the same minima, we found a simple definition:

$$f(\vec{r_i}) = \frac{1.5}{1 + \sum_{j \neq i}^{N_{points}} \delta(|r_i - r_j|)} \quad (1)$$

$$\delta(|r_i - r_j|) = \begin{cases} 1 & |r_i - r_j| < r_{max} \\ 0 & |r_i - r_j| \geq r_{max} \end{cases}$$

where $r_{max}$ is the same $r_{max}$ in the window depth scaling and $|r_i - r_j|$ is the distance between sample points i and j in the reduced parameter space. The final node selection rule used is very similar to the classic UCT or UCB with a few key modifications which is called the Upper Confidence Bound for Parameters or UCP[32].

$$UCP(\theta_i) = -\min(p_1, p_2, \dots p_{n_i}) + C \cdot f(\vec{r_i}) \cdot \sqrt{\frac{\log N_i}{n_i}} \quad (2)$$

Where $\theta_i$ represents node *i* in the MCTS structure, *p* is the reward for a given playout, C is the exploration constant, $f(\vec{r_i})$ is the uniqueness criteria value for this node, $n_i$ is the number of playout samples taken by this node and all of its child nodes, and $N_i$ is a similar value as $n_i$ except it is the parent node's playout count instead of this node's. The reward is given as the best playout reward discovered as opposed to the average since the algorithm tries to find the best solution instead of the highest probability of winning like in many other MCTS formalisms.

**Adaptive Sampling in Playouts**: In discrete space searches such as board games, playouts are performed by randomly moving pieces to evaluate game scenarios ending in a victory or a loss. In a continuous action space, there isn't a distinct "win" scenario. Rather, playouts are viewed as a request for additional random sampling around a given point. When a node is selected for a playout, we perform random vector displacements from the parameter set contained in the node. This is akin to a random walk through the phase space that is guided by the MCTS algorithm. To allow the reinforcement learning to properly determine what path to take next, it is important to ensure that the generated sample points are high in quality. There are a great many stochastical traps that one can fall into depending on the sampling method. One such problem is when generating a vector that corresponds to a perturbation of the parameter space to create a new playout. If one were to use simple distributions such as an N-dimensional uniform, gaussian, etc., where each direction is generated from its own distribution, independent of all other variables, the probability of generating a large displacement increases with the number of parameters. The probability of generating a value between ($-3\sigma$, $3\sigma$) for a 1-dimensional gaussian is ~99%. For a 100-dimensional gaussian the probability of all values being found within $3\sigma$ is $.99^{100}$ which is simply around 30%. This means the vast majority of vectors generated will have one or more extreme values. This problem becomes even more extreme as a larger number of parameters are introduced. As such better generation schemes are needed when creating points in a high-dimensional space. A simple and effective way to circumvent this is to generate a vector uniformly on the surface of an N-Sphere of radius 1 and then uniformly pick the vector length. Since we pick within a distance, R, which is a collective variable, one can show that it is actually a biased distribution.

$$\int_0^{r_{max}} \boldsymbol{dr} = \int_0^{r_{max}} J(r)f(r)\boldsymbol{dr}$$

Where $J(r)$ is the radial component of the Jacobian for the polar coordinates and $f(r)$ is the probability density function. For visual simplicity, the normalization constant is neglected in this equation. This of course assumes that the angular components have already been fixed and thus integrated out. To have a distribution that is uniform on $r$, the product of the probability density function and the Jacobian must equal a constant. This of course implies

$$f(r) = \frac{1}{J(r)}$$

If we examine the radial component of the Jacobian for an N-Sphere we find it is simply given by

$$J(r) = r^{N-1}$$

As such the probability density function regardless of the number of dimensions must equal

$$f(r) = \frac{1}{J(r)} = \frac{1}{r^{N-1}}$$

This implies the probability distribution in Cartesian space is given by

$$\int_0^{r=r_{max}} \frac{1}{(\sum_{i=1}^{N} x_i^2)^{(N-1)/2}} dx_1 dx_2 \dots dx_N$$

Thus, regardless of the number of dimensions, there will always be a reasonable probability of picking both large and small displacement vectors. This allows the reinforcement learning algorithm to determine the size of the vector needed to find a better reward function.

**Exploitation in Continuous Action Space**: To facilitate exploration in a continuous search space, we must allow the algorithm to narrow in on a solution and eventually converge. Using a constant maximum vector length is seen to find a decent solution but remains highly inefficient. Too large a step size is no better than a random search whereas too small requires several node expansions to find a good solution. Additionally, within the tree, there was little correlation between the information stored in a node and the information stored inside its parent node. In a board game MCTS algorithm, each node contains a "game state" *i.e.,* the game piece's positions on the board. A child node is related to its parent by the fact that you can obtain the child's position by moving a single piece from the parent's position. Restoring this correlation is paramount to have the MCTS algorithm formalism make any logical sense in addition to ensuring that its results are consistent.

We introduce a window scaling scheme (Fig. 2 (c)). Initially, the search space starts has bounds $[\alpha_{1,min}, \alpha_{1,max}]$ and $[\alpha_{2,min}, \alpha_{2,max}]$ respectively. And the largest vector distance $r_{max}$, corresponding to the sampling radius of the hypersphere that can be generated is given as, $r_1$. This radius is assigned to smaller and smaller values with the increasing depth of the corresponding node in the MCTS tree (Fig. 2 (c))). The reduction is done following a gaussian curve using the equation

$$r = \begin{cases} r_{max} * \exp\left(-a * \left(\frac{depth}{maxdepth}\right)^2\right), & depth \leq maxdepth \\ 0, & depth \geq maxdepth \end{cases} \quad (3)$$

"a" is the tunable parameter. The telescoping window scaling approach ensures that the algorithm is incrementally refining the phase space. This allows the algorithm to initially make larger scans of the phase space and as it finds interesting regions it is allowed to zoom in on those regions and begin exploring in more detail. Restoring the correlation between the parent and child node in that a child node is a zoomed-in region around the parent node, it gives the algorithm some direction such that the algorithm is not simply performing a purely random walk, and it also allows it to converge sufficiently close to an optimal solution since it is making smaller and smaller adjustments as it expands the tree depth.

**Data Availability:**

The dataset of metastable carbon (C) polymorphs generated using the CASTING workflow is available at https://github.com/sbanik2/CASTING. The reference structures such as unit cells of Silver (Ag -FCC), Diamond & Graphite(C), Graphane (CH), and the ground state Neodymium Nickel Oxide ($NdNiO_3$) are available in the Materials Project Database (https://materialsproject.org/).

**Code Availability:**

A pseudocode of the MCTS optimizer with an optimization code for atomic nanoclusters is available at https://github.com/sbanik2/CASTING.

**Acknowledgment**:

This work performed at the Center for Nanoscale Materials, a U.S. Department of Energy Office of Science User Facility, was supported by the U.S. DOE, Office of Basic Energy Sciences, under Contract No. DE-AC02-06CH11357. This material is based on work supported by the DOE, Office of Science, BES Data, Artificial Intelligence, and Machine Learning at DOE Scientific User Facilities programme (MLExchange). SKRS would also like to acknowledge the support from the UIC faculty start-up fund. This research used resources of the National Energy Research Scientific Computing Center (NERSC), a US Department of Energy Office of Science User Facility located at Lawrence Berkeley National Laboratory, operated under Contract No. DE-AC02-05CH11231.

**Competing Interests**

The Authors declare no Competing Financial or Non-Financial Interests.

**Author contributions**

S.B and S.K.R.S. conceived the project. S.B. developed the CASTING workflow with input from T.L. S.K.R.S., S.S. and H.C provided feedback on the CASTING workflow and the crystal structure search/validation. S.B performed all the calculations. S.B. evaluated the performance of the workflow for different representative systems and analyzed the results with input from S.K.R.S. and P.D. S.B. and S.K.R.S. wrote the manuscript with input from all co-authors. S.M and H.C provided feedback on the data analysis and the workflow hyperparameters for the target systems. All authors participated in discussing the results and provided comments and suggestions on the various sections of the manuscript. S.K.R.S supervised and directed the overall work.

# Supplementary Information

## A Continuous Action Space Tree search for INverse desiGn (CASTING) Framework for Materials Discovery

Suvo Banik[1, 2], Troy Loefller[1, 2], Sukriti Manna[1, 2], Srilok Srinivasan[1], Pierre Darancet[1], Henry Chan[1, 2], Alexander Hexemer[3], Subramanian KRS Sankaranarayanan*[1,2]

*skrssank@uic.edu

[1] Center for Nanoscale Materials, Argonne National Laboratory, Lemont, Illinois 60439.
[2] Department of Mechanical and Industrial Engineering, University of Illinois, Chicago, Illinois 60607.
[3] Advanced Light Source (ALS) Division, Lawrence Berkeley National Laboratory, Berkeley, CA 94720.

### S. 1. Calculation of the unique structures sampled by MCTS, GA, and random sampling for C polymorphs:

To calculate the unique C structures sampled during a search of any sampling algorithm (e.g., MCTS, GA, random), we follow a two-step approach. We start by calculating order parameters ($Q_2$, $Q_4$, $Q_6$)[1] + coordination vectors of the sampled structures with a cutoff of 3A°. We then use an unsupervised Agglomerative Clustering[2], with a "distance_threshold" of 1 and a "ward" linkage to cluster similar configurations in feature space. Thus, the number of clusters is the number of unique structures sampled and from each cluster, the configuration having the least energy value is chosen.

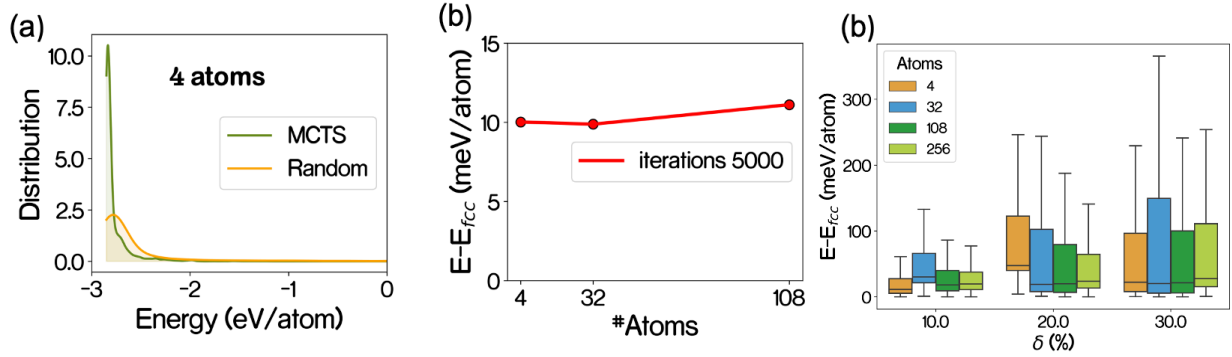

*Fig. S1.* *(a) Typical energy distribution of Ag FCC structure sampled by MCTS algorithm as compared to a purely random sampling approach for a system size of 4 atoms. (b) The effect of an increase in the number of dimensions on the best solution obtained for a bound increment ($\delta$) of 30%, using a random sampling approach. (c) The overall energy distribution of all the sampled configurations for varying dimesionalities with different lattice parameter bounds ($\delta$).*

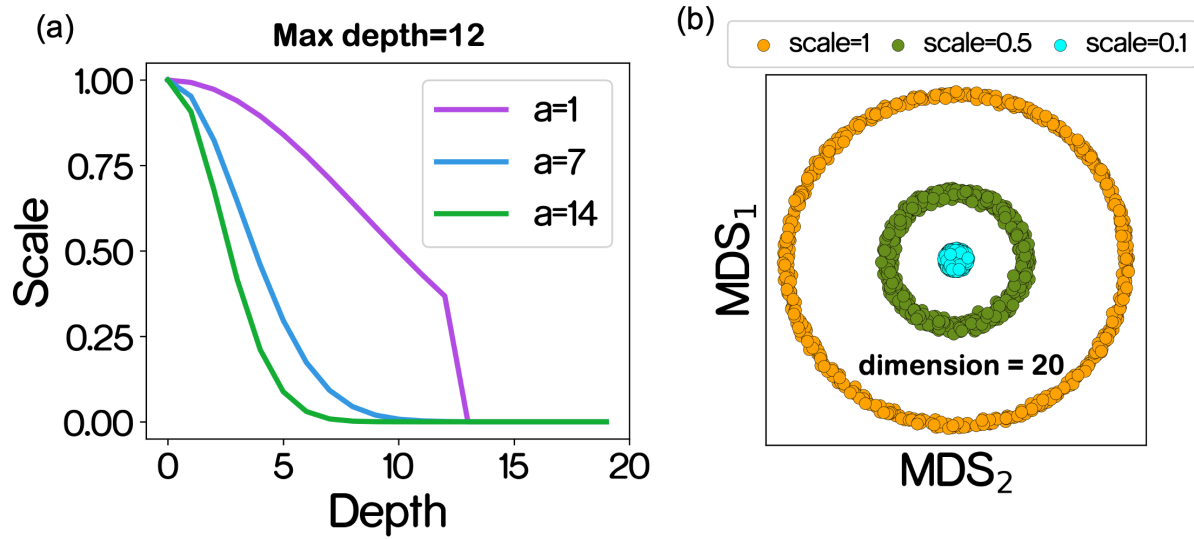

*Fig S2.* *(a) Depth scaling for different values of "a" hyperparameter. The scale reduces following Eq.2. (b) The MDS (Multi-Dimensional Scaling) plot of parameter vectors (20 dimensions) following a Hypersphere perturbation scheme for the mutation of lattice parameters and coordinates of the atoms. The points are sampled on the surface of the hypersphere, around the target point in search space. The perturbation window is scaled with depth by changing the scaling factor "a".*

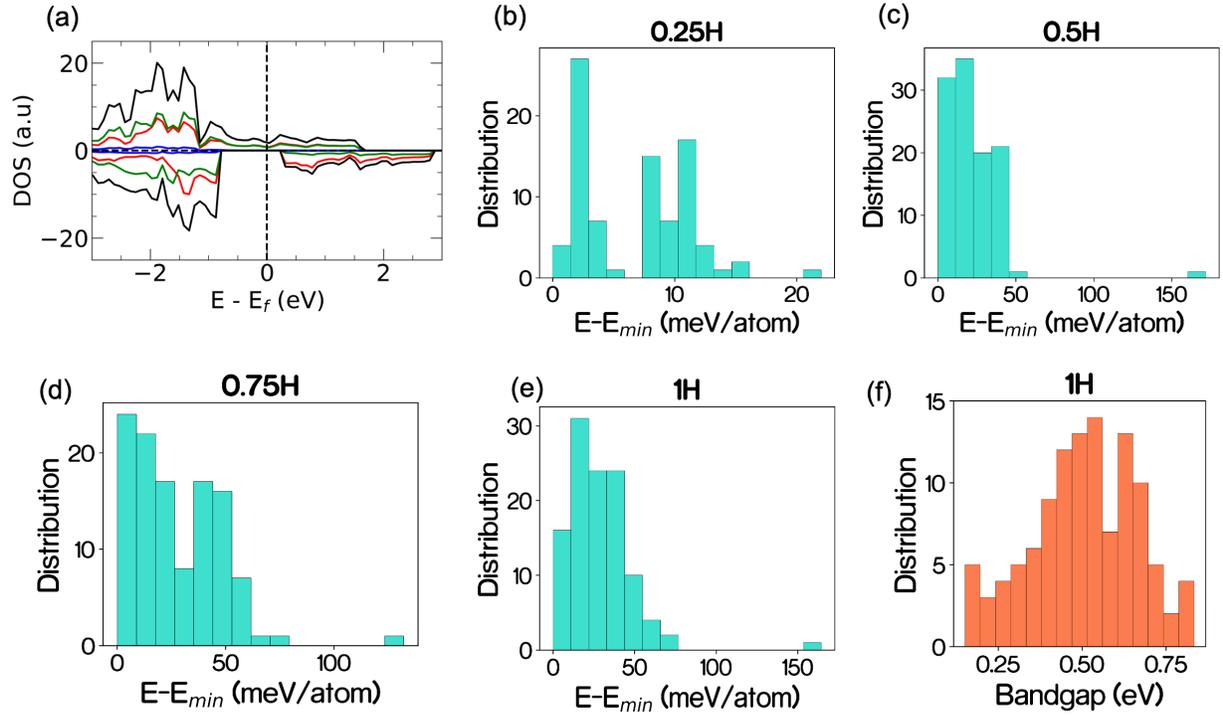

*Fig. S3. (a) The density of states (DOS) of pristine NNO. (b-e) Energy distribution of the MCTS sampled metastable configurations with a doping concentration of 0.25H, 0.5H, 0.75H, and 1H respectively. (f) Band gap distribution of the sampled configurations for a doping concentration of 1H.*

### S. 2. First-principles calculations of the electronic density of states and band gap:

The First principal calculations are computed with a DFT+U method using a VASP[3] package. A PBEsol[4] exchange-correlation functional with $U_{eff}$ = 2 eV is used to calculate the structural and electronic properties of pristine NNO and H-doped NNO. This also conforms to the calculation settings previously used to compute the pure metallic state of pure NNO and the H-induced insulating phase of H-NNO. The pseudopotentials Nd_3 (06Sep2000), Ni_pv (Ni_pv 06Sep2000), O (O 08Apr2002), and H (H 15Jun2001) are used for Nd, Ni, O, and H respectively. For the geometry optimization calculations, the plane wave cut-off energy to 500 eV, and the Brillouin zone was sampled at the Γ-point with 6 × 6 × 4 k-point mesh. For calculations of the electronic density of states and band gap, we have used a relatively dense 12 × 12 × 8 k-point mesh sampled at Γ-point. The pristine orthorhombic perovskite (space group Pbnm) NNO structures is used from the Materials Project Database[5].